\documentclass[12pt,preprint]{aastex}
\usepackage{emulateapj5}
\usepackage{graphicx}
\usepackage{grffile}
\usepackage{natbib}

\shorttitle{The Growth of Supermassive Stars and Black Holes}
\shortauthors{J.~L. Johnson, D.~J. Whalen, C.~L. Fryer, and H. Li}

\begin{document}

\title{The Growth of the Stellar Seeds of Supermassive Black Holes}

\author{Jarrett L. Johnson\altaffilmark{1}, Daniel J. Whalen\altaffilmark{2}, Chris L. Fryer\altaffilmark{3}, 
and Hui Li\altaffilmark{1}}

\affil{$^{1}$Nuclear and Particle Physics, Astrophysics and Cosmology Group (T-2), Los Alamos 
National Laboratory, Los Alamos, NM 87545}
\affil{$^{2}$McWilliams Fellow, Department of Physics, Carnegie Mellon University, Pittsburgh, PA 
15213}
\affil{$^{3}$CCS-2, Los Alamos National Laboratory, Los Alamos, NM  87545}
\email{jlj@lanl.gov}


\begin{abstract}

The collapse of baryons into extremely massive stars with masses $\ga$ 10$^4$ M$_{\odot}$ in a 
small fraction of protogalaxies at $z \ga$ 10 is a promising candidate for the origin of supermassive 
black holes, some of which grow to a billion solar masses by z $\sim$ 7.  We determine the maximum 
masses such stars can attain by accreting primordial gas.  We find that at relatively low accretion 
rates the strong ionizing radiation of these stars limits their masses to $M_{\rm *}$ $\sim$ 10$^3$ M$
_{\odot}$ ($\dot{M}_{\rm acc}$/10$^{-3}$ M$_{\odot}$ yr$^{-1}$)$^{\frac{8}{7}}$, where $\dot{M}_{\rm 
acc}$ is the rate at which the star gains mass.  However, at the higher central infall rates usually found 
in numerical simulations of protogalactic collapse ($\ga$ 0.1 M$_{\odot}$ yr$^{-1}$), the lifetime of the 
star instead limits its final mass to $\sim$ 10$^6$ M$_{\odot}$.  Furthermore, for the spherical accretion 
rates at which the star can grow, its ionizing radiation is confined deep within the protogalaxy, so the 
evolution of the star is decoupled from that of its host galaxy. Ly$\alpha$ emission from the surrounding 
H~{\sc ii} region is trapped in these heavy accretion flows and likely reprocessed into strong Balmer 
series emission, which may be observable by the {\it James Webb Space Telescope}.  This, along with 
strong He~{\sc ii} $\lambda$ 1640 and continuum emission, are likely to be the key observational 
signatures of the progenitors of supermassive black holes at high redshift.  

\end{abstract}


\section{Introduction}

The existence of 10$^8$ - 10$^9$ M$_{\odot}$ black holes (BH) in massive galaxies by $z \sim 
7$, less than a billion years after the Big Bang \citep{fan03,wil03,mort11}, remains one of the great 
mysteries of cosmological structure formation.   In the $\Lambda$CDM paradigm, early structure 
formation is hierarchical, with small dark matter halos at early epochs evolving into ever more 
massive ones by accretion and mergers through cosmic time. Hence, it is generally held that the 
supermassive black holes (SMBH) of the $z \sim 7$ \textit{Sloan Digital Sky Survey} (\textit{SDSS}) 
quasars grow from much smaller seeds at high redshifts.  The origin of these seeds, and how they 
reach such large masses by such early times, remains to be understood.
At least four main processes have 
been proposed for their formation \citep[see][]{vol10,ah11rev,primrev}: the collapse of Pop III stars into 100 - 300 M$_{\odot}$ 
BH at $z \sim 20$ \citep[e.g.][]{mad01,awa09,milos09,th09}, the direct
collapse of extremely hydrogen molecule-poor primordial gas in $\sim$ 10$^8$ M$_{\odot}$ dark 
matter halos into 10$^4$ - 10$^6$ M$_{
\odot}$ BH at $z \sim 10$ \citep[e.g.][see also Colgate et al. 2003]{bl03,kbd04,begel06,ln06,spaans06,osh08,rh09,sethi10,io11},  
the collapse of dense primeval star clusters into 10$^4$ - 10$^6$
M$_{\odot}$ BH \citep[see e.g.][]{brmvol08,dv09}, and the collapse of primordial
overdensities in the immediate aftermath of the Big Bang
\citep[see e.g.][]{mack07,carrrev}. 

The processes by which black holes form at high redshift and evolve into SMBH must account for 
how they become so large by $z \sim 7$ and why their numbers at that redshift are so small, about 
1 Gpc$^{-3}$.  Pop III seed BH are plentiful at $z \sim$ 20 - 30 but must grow at the Eddington limit 
without interruption to reach 10$^8$ - 10$^9$ M$_{\odot}$ by $z \sim$ 7.  This is problematic 
because they and their progenitors either expel all the baryons from the shallow potential wells of 
the halos that create them, so they are "born starving" \citep{wan04,jb07,pdc07,awa09,jeon11}, or 
they eject themselves from their halos, and thus their fuel supply, at hundreds of km/s if they are born 
in core-collapse supernova explosions (Whalen \& Fryer 2011 in prep).  Also, accretion onto Pop III 
BH has been found to be inefficient on small scales, typically at most 20\% Eddington \citep{milos09,
pm11a}, making the constant duty cycles required for sustained growth difficult \citep[but see][]{Li11}.  

If halos can instead congregate into primitive galaxies of $\sim$ 10$^8$ M$_{\odot}$ at $z \sim$ 10 - 15 
that are devoid of the coolant molecular hydrogen (H$_{\rm 2}$), they reach virial temperatures of $\sim$ 10$^4$ K and begin to 
atomically cool.  Numerical simulations of this process \citep{wta08,rh09,sbh10} find that in analogy to 
Pop III star formation in much smaller halos at higher redshifts \citep{nu01,bcl02,abn02,on07,y08,turk09,
stacy10,clark11,greif11}, baryons rapidly pool at the center of the halo and form a hydrostatic object. 
But this object is thought to become far more massive than Pop III stars modeled to date because atomic 
line cooling and the deeper potential well of the halo lead to much higher infall rates at its center, 0.1 - 1 
M$_{\odot}$ yr$^{-1}$ rather than the 10$^{-4}$ M$_{\odot}$ yr$^{-1}$ typical of primordial star-forming 
minihalos at $z \sim$ 20.  Such objects could collapse into black holes that are far more massive than 
Pop III BH \citep[e.g.][]{shib02}, with Bondi-Hoyle accretion rates that allow them to grow into SMBH in 
less time \citep[e.g. Wyithe \& Loeb 2011; but see][]{dotan2011}.

One difficulty with this scenario is that it is not yet understood how primitive galaxies can form in the high 
Lyman-Werner UV backgrounds needed to fully suppress H$_{\rm 2}$
molecule formation and quench Pop III star formation in its constituent halos prior 
to assembly \citep[1 - 10 times those expected at the epoch of reionization at $z \sim 6$;][see also Agarwal et al. in prep]
{dijkstra08,sbh10,whb11}.  
Furthermore, it is not known if the object at the center of the protogalaxy even becomes a star, because at 
such high accretion rates it reaches very large masses on timescales that are short in comparison to 
Kelvin-Helmholtz times and the onset of nuclear burning (although Ohkubo et al. 2009 have modeled the 
evolution of Pop III stars under accretion at much lower rates).  It could be that the central object reaches 
such high entropies and densities that it is enveloped by an event horizon before reaching the main 
sequence \citep{fwh01}.  If the central object becomes a star, what governs its rate of growth, its final 
mass, and thus the mass of the SMBH seed it becomes?  
Numerous authors have studied how stellar radiation regulates accretion onto Pop III protostars, but at much lower inflow 
rates than those in the centers of collapsing protogalaxies \citep[e.g.][]{stahler86b,op01,oi02,tm08,stacy11,hoyy11}.  Do the much higher accretion rates in  
primitive galaxies quench radiative feedback?   

We have performed extensive semianalytical calculations of radiative feedback by Pop III supermassive
stars in collapsing protogalaxies at $z \sim$ 10.  In $\S \, 2$, we examine the primary forms of radiative
feedback on accretion onto the star (deferring the processes that can be ignored to thorough examination
in the Appendix).  We also derive the maximum mass that the supermassive star, and hence the SMBH 
seed, can achieve as a function of accretion rate, accounting for its prodigious ionizing UV flux.  In $\S \, 3$, 
we estimate the maximum mass that the star can reach if it is limited by its finite lifetime rather than by 
radiative feedback.  We discuss the observational signatures of rapidly accreting supermassive primordial 
stars in $\S \, 4$.  Finally, in $\S \, 5$ we review the implications of final supermassive stellar mass for SMBH 
seed mass and the appearance of the first quasars in the Universe.

\begin{figure*}[t]
  \centering
  \includegraphics[width=3.5in]{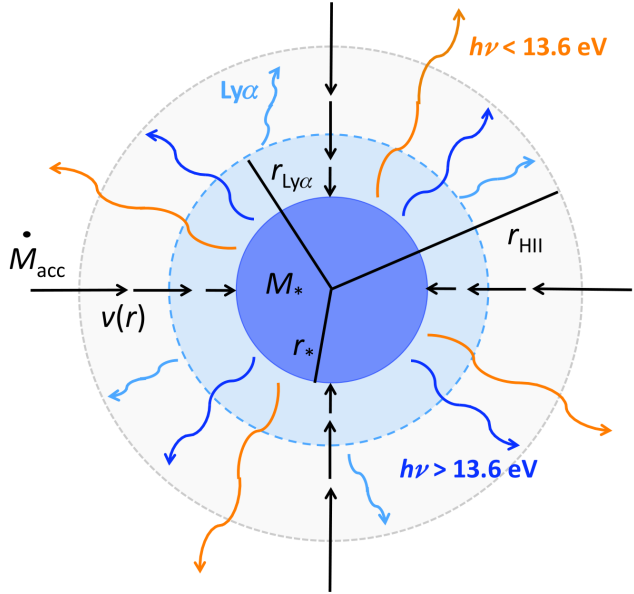}
  \caption{
Schematic representation of accretion onto a supermassive Pop~III star of mass $M_{\rm *}$ and 
radius $r_{\rm *}$.  Gas accretes at a constant rate $\dot{M}_{\rm acc}$, being acted upon only by 
gravity at $r$ $>$ $r_{\rm HII}$, the radius of the H~{\sc ii} region of the star.  Inside $r_{\rm HII}$ 
the gas, whose infall velocity is $v(r)$, is decelerated by momentum absorbed during ionizations, 
which is indicated by the shortening of the velocity vector arrows.  While continuum photons with 
energies $h\nu$ $<$ 13.6 eV escape the H~{\sc ii} region and do not couple strongly to the gas 
even outside $r_{\rm HII}$, resonant line photons do couple strongly and so in principle can also
impart momentum to the gas.  In particular, a large amount of momentum is emitted by H~{\sc i} 
recombinations as Ly$\alpha$; however, these photons couple so strongly to the gas that within 
$r_{\rm Ly\alpha}$ more than 50 percent of the emission is trapped in the accretion flow and 
cannot propagate outward and affect the dynamics of the flow.  Within $r_{\rm HII}$, Ly$\alpha$ 
photons are reprocessed into Balmer series photons which escape the H~{\sc ii} region and may 
provide an observable signature of accreting supermassive stars (see $\S \, 4$).}
\end{figure*}

\section{Radiative feedback-limited accretion}

We adopt an analytical approach to estimate the maximum mass that an accreting primordial star 
can ultimately attain in a collapsing 10$^7$ - 10$^8$ M$_{\odot}$ protogalaxy.  For simplicity, we 
assume that accretion onto the star is constant and spherically-symmetric.  We first consider the 
ionizing UV radiation emitted by the star, which is the dominant form of radiative feedback on the 
infall and envelops the star with an H~{\sc ii} region.  In particular, we make the following 
assumptions on the nature of the radiative feedback on the accretion flow (see Fig. 1):
\vspace{0.1in}

(1) Outside the H~{\sc ii} region, gas is accelerated only by gravity and falls inwards unchecked 
by gas or radiation pressure.

\vspace{0.1in}

(2) Within the H~{\sc ii} region, the gas is decelerated due to photoionization pressure only.  This 
follows from the fact that the massive stars we consider here radiate at very nearly the Eddington 
limit.  By definition, this implies that the force due to electron scattering in the H~{\sc ii} region 
exactly cancels that due to gravity; therefore, for simplicity we can assume that the only net force 
on the gas is due to photoionization pressure \citep[see also][]{oi02}.   

In this Section we first determine how ionizing radiation from the supermassive star limits its growth 
via accretion.  We then consider the effect of the Ly$\alpha$ photons to which the majority of the 
energy in the ionizing radiation is converted.  We examine and exclude other processes that can 
contribute to feedback on accretion in the Appendix. 

\subsection{Suppression of Accretion by Photoionization}

As we are considering constant, spherically-symmetric accretion, the equation that links the number 
density $n$ of hydrogen nuclei and the velocity $v$ of the gas at a distance $r$ from the star is:
\begin{equation}
n(r) = \frac{\dot{M}_{\rm acc}}{4 \pi r^2 v(r) \mu m_{\rm H}} \mbox{\ .} \vspace{0.1in}
\end{equation}  
Here, we assume that the gas is primordial with an average atomic weight of $\mu$$m_{\rm H}$ = 
1.2$m_{\rm H}$, where $m_{\rm H}$ is the mass of the hydrogen atom.

Because the ionizing photons are trapped within the H~{\sc ii} region, we assume that the velocity of 
the gas at its edge, at a distance $r_{\rm HII}$, is just the free-fall velocity: \vspace{0.075in}
\begin{equation}
v(r_{\rm HII}) =  - \left(\frac{2GM_{\rm *}}{r_{\rm HII}}\right)^{\frac{1}{2}} \mbox{\ .} \vspace{0.075in}
\end{equation}
We note that this velocity is much higher than the sound speed of the gas, at which infall is found to 
proceed from large radii in cosmological simulations \citep{wta08,sbh10,jlj11}, so we neglect this 
small contribution to the gas velocity for simplicity.

The radius of the H~{\sc ii} region, $r_{\rm HII}$, is determined by the ionization front jump conditions, 
which require a balance between the rate $Q$ at which ionizing photons are emitted from the star and 
the sum of the rate of hydrogen recombinations in the ionized volume (first term on the RHS in equation 
3) and the rate at which neutral atoms enter the H~{\sc ii} region through the accretion flow (second term 
on the RHS in equation 3).  Following \citet{raiter10}, we note that the rate at which ionizations occur is 
higher than $Q$ by a factor $P$ which accounts for corrections to the neutral hydrogen level populations 
at the high temperatures \citep[$\simeq$  2-4 $\times$ 10$^4$ K; e.g.][]{wan04,abs06} expected for the H
{\sc ii} regions of massive Pop~III stars.  To account for this rate, we take it that the total ionization rate in 
the H {\sc ii} region is $Q_{\rm eff}$ = $P$$Q$, where $P$ is the ratio of the average ionizing photon 
energy and the ionization potential for neutral hydrogen, 13.6 eV.  Integrating over the density profile of 
the gas in eq. 1 to find the radius at which the total number of recombinations in the enclosed volume 
balances the total number of photoionizations, we arrive at the following expression for $r_{\rm HII}$:
\begin{eqnarray}
Q_{\rm eff} & \simeq & \int^{r_{\rm HII}}_{r_{\rm *}} 4 \pi \alpha_{\rm B} r^2 n^2 dr + \frac{\dot{M}_{\rm acc}}{\mu m_{\rm H}}      \nonumber \\
& = & \int^{r_{\rm HII}}_{r_{\rm *}} 4 \pi \alpha_{\rm B} r^2 \left[\frac{\dot{M}_{\rm acc}}{4 \pi r^2 v(r) \mu m_{\rm H}} \right]^2 dr + \frac{\dot{M}_{\rm acc}}{\mu m_{\rm H}}\mbox{\ ,} 
\end{eqnarray}
where $\alpha_{\rm B}$ $\simeq$ 1.4 $\times$ 10$^{-13}$ cm$^{3}$ s$^{-1}$ is the recombination 
coefficient for $\ga$ 2 $\times$ 10$^4$ K photoionized primordial gas \citep{wan04,osterbrock06}.  
Following \citet{oi02}, for simplicity we neglect the second term in our calculations, as it is in general 
much smaller than $Q_{\rm eff}$.  This is true in particular for the solutions that we find for the 
minimum accretion rate onto a star of a given mass; at much higher rates, this term may become 
large, likely limiting the radius of the H~{\sc ii} region.

Finally, once the accreting gas crosses the stationary ionization front (as shown in Fig. 1), the equation 
that governs its deceleration due to momentum imparted by photoionizations is:
\begin{equation}
v \frac{dv}{dr} = -\frac{\alpha_{\rm B}  h \nu n }{\mu m_{\rm H} c} = -\frac{\alpha_{\rm B} h \nu}{\mu m_{\rm H} c} \left[\frac{\dot{M}_{\rm acc}}{4 \pi r^2 v(r) \mu m_{\rm H}} \right] \mbox{\ ,}
\end{equation} 
where again we have assumed that the photoionization rate of hydrogen is balanced its recombination
rate, $\alpha_{\rm B}n$, and that the momentum transferred to the gas
per photoionization is $h\nu/c$.  We note again that here we have not
included the force due to gravity or that due to Thomson scattering,
as these two forces cancel one another since the star is assumed to
radiate at the Eddington rate.
For the massive primordial stars that we consider here, we take the effective surface temperature of the 
star to always be $\simeq$ 10$^5$ K, which yields an average energy per ionizing photon $h\nu$ 
$\simeq$ 29 eV.  In turn, this implies $P$ $\equiv$ $h\nu$/13.6 eV = 2.1, which we assume throughout 
our study.  In equation (4) we again choose case B, as we assume that recombinations to the ground 
level of hydrogen result in the emission of photons which do not deposit a net momentum to the gas 
(being emitted isotropically and contributing to the diffuse radiation field).\footnote{At the high densities 
in the H~{\sc ii} regions we consider (Fig. 2) and with the enhanced population of neutral hydrogen in the 
$n$ = 2 state due high temperatures in primordial H~{\sc ii} regions \citep{raiter10}, a $\sim$ 30 percent 
smaller 'extended case B' \citep{hs87} recombination coefficient may be more appropriate, but our final 
results are not strongly dependent on this choice.}

\begin{figure*}[t]
  \centering
  \includegraphics[width=4.75in]{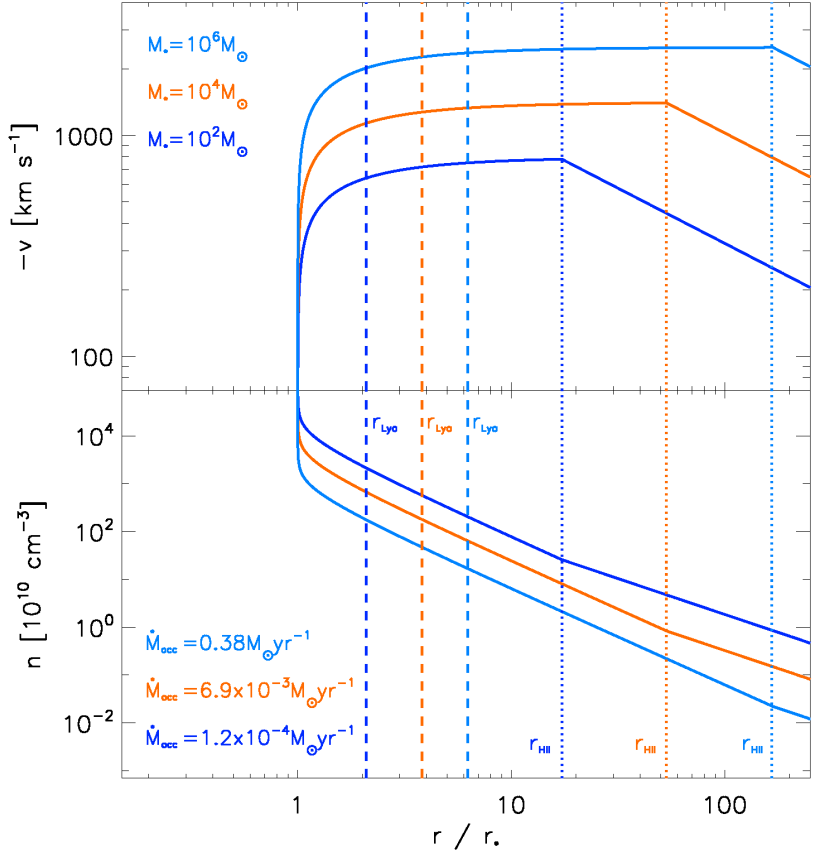}
  \caption{The infall velocity ({\it top panel}) and density ({\it bottom panel}) of the gas accreting onto 
primordial stars as a function of distance $r$ from the center of the star for three stellar masses: 10$
^2$ ({\it dark blue}), 10$^4$ ({\it red}), and 10$^6$ M$_{\odot}$ ({\it light blue}).  Each star accretes 
at the minimum rate (as labeled) its strong ionizing flux permits, as given by equation 11. The dotted 
lines show the radius $r_{\rm HII}$ of the H~{\sc ii} regions surrounding the stars, with their colors 
corresponding to those of the respective stellar masses.  Outside the H~{\sc ii} region, the gas is in 
free-fall, but after crossing into the H~{\sc ii} region boundary it is rapidly photoionized and 
decelerated until it arrives at the stellar surface with velocity $v(r_{\rm *})$ = 0. The dashed lines 
show the trapping radius $r_{\rm Ly\alpha}$ for Ly$\alpha$ photons from recombinations in the H II 
region (assuming a gas temperature of 4 $\times$ 10$^4$ K, see equation 22); partly because the 
vast majority of these photons are trapped within $r_{\rm Ly\alpha}$, we can neglect Ly$\alpha$ 
scattering feedback as discussed in $\S \, 2.2$.  Accretion at rates below those shown here is 
not possible, as in such cases $r_{\rm HII}$ $\to$ $\infty$ and photoionization pressure halts infall at 
all radii.}

\end{figure*}

To find the solutions for steady state accretion, we must solve the equation of motion equation 4.  
In particular, to find the minimum accretion rate for which an inflow solution exists, we search for 
solutions for which the infall velocity of the gas goes to zero at the stellar surface $r_{\rm *}$:
\vspace{0.05in}
\begin{equation}
v(r_{\rm *}) = 0 \mbox{\ .} \vspace{0.05in}
\end{equation}  
That is, we must simultaneously solve equations 3 and 4, under the constraints given by equations 
2 and 5.  These constraints yield the maximum mass that the star can achieve by steady accretion 
before its intense ionizing radiation halts infall at $r$ $>$ $r_{\rm *}$ and prevents its further growth.  
On the other hand, they also yield the lower bound for the final mass of the central object, since for 
$v(r_{\rm *}) > 0$ radiative feedback fails to cut off accretion and the star will grow even faster than 
when gas merely comes to a halt on its surface.  

The solution to equation 4 has the general form \vspace{0.075in}
\begin{equation}
v(r) = \left[\left(\frac{3\alpha_{\rm B}h\nu\dot{M}_{\rm acc}}{4 \pi \mu^2 m_{\rm H}^2 c}\right)r^{-1} - K \right]^{\frac{1}{3}} \mbox{\ ,} \vspace{0.075in}
\end{equation}
where $K$ is a constant whose value must satisfy the constraint that $v(r_{\rm *})$ = 0.  This yields \vspace{0.075in}
for $K$ 
\begin{equation}
K = \left(\frac{3 \alpha_{\rm B} h \nu \dot{M}_{\rm acc}}{4 \pi \mu^2 m_{\rm H}^2 c }\right)r_{\rm *}^{-1}. \vspace{0.075in}   
\end{equation}
The equation for the infall velocity in the H II region then becomes \vspace{0.075in}
\begin{equation}
v(r) = \left[ \left(\frac{3 \alpha_{\rm B} h \nu \dot{M}_{\rm acc}}{4 \pi \mu^2 m_{\rm H}^2 c }\right)r^{-1}  -\left(\frac{3\alpha_{\rm B}h\nu\dot{M}_{\rm acc}}{4 \pi \mu^2 m_{\rm H}^2 c}\right)r_{\rm *}^{-1} \right]^{\frac{1}{3}} \mbox {\ .}
\end{equation}
As shown in Figure 2, these solutions give a relatively constant velocity for the gas in the H II region until 
it is very close to the stellar surface.  In turn, equation (1) implies that $n(r)$ is essentially $\propto$ $r^{-2}$ 
in the H~{\sc ii} region.  

Applying this $v(r)$ (and hence $n(r)$) to the integral in equation 3 and evaluating it yields $r_{\rm HII}$:
\begin{equation}
r_{\rm HII} =  \left[r_{\rm *}^{-1}  - \frac{4 \pi Q_{\rm eff}^3}{3 \alpha_{\rm B}} \left(\frac{h \nu \mu m_{\rm H} }{c} \right)^2  \dot{M}_{\rm acc}^{-4}  \right]^{-1} \mbox{\ .} \vspace{0.1in}
\end{equation}

Finally, we must satisfy the constraint given by equation 2 to find the solution for the minimum 
accretion rate.  Substituting equations 8 and 9 into equation 2 and rearranging the terms gives
the following quadratic equation in $\dot{M^2_{\rm acc}}$:

\begin{eqnarray}
0 & = & \left(\frac{2 G M_{\rm *}}{r_{\rm *}} \right)\dot{M^4_{\rm acc}} - \left(\frac{Q_{\rm eff} h \nu}{c} \right)^2 \dot{M^2_{\rm acc}} \nonumber \\
& - & \frac{8 \pi G M_{\rm *} Q_{\rm eff}^3}{3 \alpha_{\rm B}} \left(\frac{h \nu \mu m_{\rm H}}{c} \right)^2 \mbox{\ ,}
\end{eqnarray}
whose solution is

\begin{equation}
\dot{M^2_{\rm acc}} = \frac{\left(\frac{Q_{\rm eff} h \nu}{c}\right)^2 + \left[\left(\frac{Q_{\rm eff} h \nu}{c}\right)^4 + \frac{64 \pi  Q_{\rm eff}^3}{3 \alpha_{\rm B} r_{\rm *}} \left(\frac{ G M_{\rm *} h \nu \mu m_{\rm H}}{c} \right)^2  \right]^{\frac{1}{2}}}{\frac{4GM_{\rm *}}{r_{\rm *}}} \mbox{\ .}
\end{equation}
Noting that the second term in the square brackets is much larger than the first, we have 
\begin{equation}
\dot{M}_{\rm acc}\simeq \left(\frac{4 \pi  Q_{\rm eff}^3 r_{\rm *}}{3 \alpha_{\rm B}} \left(\frac{h \nu \mu m_{\rm H}}{c}\right)^2  \right)^{\frac{1}{4}} \mbox{\ ,}
\end{equation}
which exactly matches the accretion rate for which the H~{\sc ii} region breaks out to infinity (see 
equation 9) and therefore stops accretion entirely because photoionization pressure exerts a net 
outward force on gas at all radii.\footnote{While in our solution the H~{\sc ii} region formally extends 
to infinity in this case, for accretion to be terminated it need only extend to the Bondi radius, outside 
of which gas cannot be gravitationally captured by the star.}  Thus, we see that the H~{\sc ii} region 
is confined and that accretion proceeds at infall rates higher than those given by equation 12 and 
that no inflow solutions exist for lower rates.\footnote{Relatively small deviations from the solutions 
we have found for $v(r)$ at $r_{\rm HII}$ may result in a shock developing there.  However, even for 
a strong shock for which there is a density jump of a factor of four, the minimum accretion rates in equation 11
change by $\la$ 20 percent.}  

Now, by expressing $Q$ ( where again $Q_{\rm eff}$ = $P$$Q$, with $P$ = 2.1) and $r_{\rm *}$ just 
in terms of the stellar mass $M_{\rm *}$, we can find the maximum stellar mass for which accretion 
onto the star is permitted (at $\dot{M}_{\rm acc}$ in equation 12).  We can take it that, if the star is 
thermally relaxed\footnote{At the minimum accretion rates we find for
  a given stellar mass (equation 15), the 'trapping' 
radius due to electron scattering \citep[which is proportional to the accretion rate, e.g.][]{begel78} is 
always smaller than the stellar radius given by equation 14. Therefore, the main sequence stars we 
consider here are thermally relaxed \citep[see also][]{ohk09}. At higher accretion rates, the radiation 
emitted by the star can become trapped due to electron scattering, resulting in an expansion of the 
stellar photosphere.  In this case the star emits fewer ionizing photons than given by
equation (13) and the radiative feedback is thus weakened, as we discuss briefly in the appendix.} 
and radiating at the Eddington limit \citep{begel10}, then the following two equations relate the mass 
$M_{\rm *}$ of the star to its ionizing photon emission rate $Q$ and radius $r_{\rm *}$: \vspace{0.05in}
\begin{equation}
Q = 1.6 \times 10^{50} \, {\rm s}^{-1} \left(\frac{M_{\rm *}}{100 \, {\rm M}_{\odot}} \right) \mbox {\ ,} \vspace{0.1in}
\end{equation}
and
\begin{equation}
r_{\rm *} = 3.7 \, {\rm R}_{\odot} \left(\frac{M_{\rm *}}{100 \, {\rm M}_{\odot}} \right)^{\frac{1}{2}} \mbox{\ .} 
\end{equation}
These are from Table 1 of \citet{bkl01} and are in good agreement with \citet{s02} and \citet{begel10}.  
With these expressions, we find that the solution to equation 12
is 
\begin{equation}
M_{\rm *} \simeq 10^3 \, {\rm M}_{\odot} \, \left(\frac{\dot{M}_{\rm acc}}{10^{-3} \, {\rm M}_{\odot} \, {\rm yr}^{-1}}\right)^{\frac{8}{7}},
\end{equation}
where $M_{\rm *}$ is the maximum stellar mass attainable under accretion of gas at a rate $\dot{M}_{\rm acc}$.  

We plot this maximum stellar mass in Figure 3.  As we shall see, the feedback due to ionizing 
radiation emitted by the star can limit its mass at relatively low accretion rates, which we term 
'feedback-limited accretion'. However, at higher accretion rates the finite lifetime of the star 
governs its maximum mass; we term this 'time-limited accretion'.  We discuss why this is so in 
$\S \, 3$. 

Although we have focused on the pressure due to hydrogen photoionizations, it is straightforward 
to include He~{\sc i} photoionizations under the condition that within the H~{\sc ii} region He~{\sc i} 
is also photoionized to He~{\sc ii}.   For the relatively hard spectra of massive primordial stars this 
is a sound assumption because the number of He~{\sc i}-ionizing photons is comparable to the 
number of H~{\sc i}-ionizing photons \citep{s02}.  The ratio of the radiation pressures from helium 
and hydrogen photoionization is then 
\begin{equation}
\frac{p_{\rm HeI}}{p_{\rm HI}} \simeq \frac{n_{\rm He}}{n_{\rm H}} \frac{h\nu_{\rm HeI}}{h\nu_{\rm HI}} \frac{\alpha_{\rm B,HeI}}{\alpha_{\rm B,HI}} \simeq 0.14 \mbox{\ ,}
\end{equation}
where the first term is the ratio of helium and hydrogen number densities, which is $\simeq 0.1$. 
The second term is the ratio of the average ionizing photon energies for He~{\sc i} and H~{\sc i}, 
which by integrating the stellar spectrum of a 10$^5$ K primordial star we find to be 38 eV/29 eV 
$\simeq$ 1.3.  The third term is the ratio of He~{\sc i} and H~{\sc i} case B recombination 
coefficients, which is $\simeq$ 1.1 \citep{osterbrock06}.  While this is a relatively modest 
increase in the radiation pressure, for completeness we include it as a multiplicative coefficient 
when solving equation 4.  Other forms of radiation pressure are not so easily accommodated by 
our analytical approach but can be neglected without strongly affecting our conclusions, as we 
discuss in the next Section and in the Appendix.

\subsection{Trapping of Ly$\alpha$ photons}

In the H~{\sc ii} region of the accreting star, a large fraction of the ionizing radiation is reprocessed 
into Lyman $\alpha$ photons, which couple very strongly to primordial gas and could impart 
significant momentum to it, as they in principle can scatter many times across the H~{\sc ii} region 
\citep[e.g.][]{Doro,oh02,tm08}.  However, while these authors showed
that scattering Ly$\alpha$ can strongly counteract the accretion flow
at relatively low accretion rates, at the extremely high rates under
which supermassive stars grow (e.g. $\dot{M_{\rm acc}}$ $\ga$
10$^{-2}$ M$_{\odot}$ yr$^{-1}$) the impact on the accretion flow is
reduced due both to the large momentum of the infalling gas and to
the trapping of Ly$\alpha$ photons within the H~{\sc ii} region.   

To show this, we first compare the momentum in Ly$\alpha$ photons to the momentum of the accretion flow.  The former is given by \citep{raiter10}
\begin{equation}
\frac{L_{\rm Ly\alpha} }{c} \simeq \beta_{\rm Ly\alpha} h \nu_{\rm Ly\alpha} \frac{Q_{\rm eff}}{c} \simeq 10^{29} \, {\rm g \, cm \, s}^{-2} \left(\frac{M_{\rm *}}{100 \, {\rm M}_{\odot}} \right)\mbox{\ ,}
\end{equation}
where $L_{\rm Ly\alpha}$ is the Ly$\alpha$ luminosity due to recombinations in the H~{\sc ii} 
region, in which $\beta_{\rm Ly\alpha}$ ($\simeq$ 0.9) Ly$\alpha$ photons, each with an energy 
$h\nu_{\rm Ly\alpha}$ = 1.6 $\times$ 10$^{-11}$ erg, are emitted for every recombination \citep{
osterbrock06,raiter10}.\footnote{We note that in the high densities in the H~{\sc ii} regions we 
consider (see Fig. 2), frequent collisions can prevent the radiative decay of hydrogen via two 
photon emission, which raises the Ly$\alpha$ luminosity above that expected in the low density 
regime, where  $\beta_{\rm Ly\alpha}$ $\simeq$ 0.68 \citep{spitzer78}.}  Here, we have also 
related the number of ionizing photons emitted per second $Q$ to the stellar mass $M_{\rm *}$ 
with equation 13.  The momentum of the accretion flow is
\begin{equation}
\dot{M}_{\rm acc} v \simeq 6 \times 10^{33} \, {\rm g \, cm \, s}^{-2} \left(\frac{v}{10^3 \, {\rm km} \, {\rm s}^{-1}} \right) \left(\frac{\dot{M}_{\rm acc}}{{\rm M}_{\odot} \, {\rm yr}^{-1}} \right)\mbox{\ ,}
\end{equation}
where we have normalized the infall velocity $v$ to its typical value at the edge of the H~{\sc ii} 
region in the solutions shown in Fig. 2, since this is where Lyman $\alpha$ photons will be most 
strongly coupled to the gas (indeed, these photons are trapped within the accretion flow at $r_{
\rm HII}$, as shown below).  Equating these two expressions and using the relation between $
\dot{M}_{\rm acc}$ and $M_{\rm *,max}$ in equation 15, we find the conditions under which 
Lyman $\alpha$ radiation pressure could halt accretion:   
\begin{equation}
v \la 600 \, {\rm km} \, {\rm s}^{-1} \, \left(\frac{\dot{M}_{\rm acc}}{0.1 \, {\rm M}_{\odot} \, {\rm yr}^{-1}}\right)^{\frac{1}{7}} \mbox{\ .}
\end{equation}
Since the inflow velocity near the edge of the H~{\sc ii} region is well above this value at the 
typical minimum accretion rates we find ($\la$ 0.1 M$_{\odot}$ yr$^{-1}$; see Fig. 2), it is clear
that Lyman $\alpha$ radiation pressure has only a small impact on accretion in comparison to 
photoionization pressure.  In particular, over the range of stellar masses and critical accretion 
rates that we consider (Figs. 2 and 3), we find that the infall velocity is $\sim$ three times that 
in equation 19.  In other words, at the high accretion rates we find, the momentum of inflow is 
always roughly three times what could be countered by Ly$\alpha$ scattering.  

It might be thought that if a Lyman $\alpha$ photon is scattered and then traverses the H~{\sc 
ii} region, it could subsequently be scattered many times, thereby enhancing the momentum it 
imparts the gas \citep[see][]{adams72,milos09}.  However, upon scattering from an atom in the 
rapid ($\simeq$ 10$^3$ km s$^{-1}$) accretion flow, the photon would be strongly blue-shifted 
relative to the gas entering the opposite side of the H~{\sc ii} region; as a result, it would couple 
to gas much less strongly thereafter, greatly limiting the degree to which it could add more 
momentum to the gas \citep[see][on how galactic outflows enhance Ly$\alpha$ escape by this 
process]{dw10}.

Furthermore, another effect which dramatically lessens the impact of Ly$\alpha$ photons on 
accretion is that their large resonant scattering cross section ensures that they are trapped in 
the H~{\sc ii} region and do not scatter out to larger radii.  To see
this, we follow the formula given by \citet{
begel78} for the trapping radius due to
Thomson scattering, $r_{\rm trap}$ $\equiv$ $\dot{M_{\rm acc}}$$\sigma_{\rm
  T}$/4$\pi$$m_{\rm H}$$c$.  We define the trapping radius for Ly$\alpha$ photons
$r_{\rm Ly\alpha}$ by substituting the cross section for Lyman
$\alpha$ scattering $\sigma_{\rm Ly\alpha}$  for the Thomson cross section $\sigma_{\rm T}$ for
electron scattering, and by accounting for the fraction of neutral
hydrogen atoms off of which Ly$\alpha$ photons can scatter.  This yields

\begin{equation}
r_{\rm Ly\alpha} \simeq 5 \times 10^{24} \, {\rm cm} \, \left(\frac{\dot{M}_{\rm acc}}{{\rm M}_{\odot} \, {\rm yr}^{-1}} \right) f_{\rm HI} \left(\frac{T}{10^4 \, {\rm K}} \right)^{-\frac{1}{2}} \mbox{\ ,}
\end{equation}
where $f_{\rm HI}$ is the neutral hydrogen fraction, $T$ is the temperature of the ionized gas, and 
we have used $\sigma_{\rm Ly\alpha} =$ 5.9 $\times$ 10$^{-14}$ cm$^{2}$ 
($T$/10$^4$K)$^{-\frac{1}{2}}$ \citep[e.g.][]{milos09}.  
The fraction of neutral hydrogen in the H~{\sc ii} region can be estimated by assuming photoionization 
equilibrium\footnote{Comparing the rates of hydrogen photoionization and collisional ionization
using the rate given by \citet{cen92}, we find that the former is much greater than the latter; thus, 
photoionization equilibrium is a valid assumption.}; we obtain 
\begin{eqnarray}
f_{\rm HI} & \simeq &\frac{4 \pi \alpha_{\rm B} n(r) r^2}{Q_{\rm eff} \sigma_{\rm HI}} \nonumber \\
& = & 3 \times 10^{-5} \, \left(\frac{n}{10^{10} \, {\rm cm}^{-3}}\right)  \left(\frac{r}{10^{15} \, {\rm cm}} \right)^2 \left(\frac{Q_{\rm eff}}{10^{50} \, {\rm s}^{-1}} \right)^{-1}     \mbox{\ ,}
\end{eqnarray}
where $\sigma_{\rm HI}$ = 6 $\times$ 10$^{-18}$ cm$^2$ is the photoionization cross section 
for H~{\sc i} \citep[although it is likely slightly lower due to the relatively hard spectrum of the 
star; see e.g.][]{jkSN11}. Then with this $f_{\rm HI}$, $Q$ from equation 13, $n(r)$ from equation 
1, and $\dot{M}_{\rm acc}$ in terms of $M_{\rm *}$ from equation 15, we find:
\begin{eqnarray}
r_{\rm Ly\alpha} & \simeq &  10^{12} \, {\rm cm} \, \left(\frac{M_{\rm *}}{100 \, {\rm M}_{\odot}} \right)^{\frac{3}{4}}   \nonumber \\
& \times & \left(\frac{v}{10^3 \, {\rm km} \, {\rm s}^{-1}} \right)^{-1} \left(\frac{T}{10^4 \, {\rm K}} \right)^{-\frac{1}{2}}  \mbox{\ .}
\end{eqnarray}
Assuming a temperature of 4 $\times$ 10$^4$ K for the photoionized primordial gas \citep{
wan04,ket04}, Ly$\alpha$ trapping radii for three cases are shown in Fig. 2.  We see that $r
_{\rm Ly\alpha}$ is $\sim$ 2 - 6 $r_{\rm *}$ and is smaller than $r_{\rm HII}$ by roughly an 
order of magnitude.  While this is a modest fraction of the volume of the H~{\sc ii} region, 
because of its steep density profile ($n$ $\propto$ r$^{-2}$) most of the Ly$\alpha$ photons 
originate from this region.  

Integration of equation 3 shows that the fraction of recombinations produced in the H~{\sc ii} 
region as a function of $r$ is $\sim$ ($r_{\rm *}^{-1}$ - $r^{-1}$)/($r_{\rm *}^{-1}$ - $r_{\rm HII}^
{-1}$).  Consequently, the fraction of Ly$\alpha$ photons originating from within the trapping 
radius is ($r_{\rm *}^{-1}$ - $r_{\rm Ly\alpha}^{-1}$)/($r_{\rm *}^{-1}$ - $r_{\rm HII}^{-1}$), or 
$\ga$ 0.5.  Thus, most of them are trapped deep in the H~{\sc ii} region and cannot propagate 
outward to slow accretion at greater radii.  Ly$\alpha$ scattering thus removes only $\sim$ 15\% 
of the momentum of infall, not 30\%, so we are justified in neglecting its impact on accretion flow.


At this point, we can show that Ly$\alpha$ photons created outside $r_{\rm Ly\alpha}$ but within 
$r_{\rm HII}$ are also trapped in the accretion flow at the boundary of the H~{\sc ii} region due to 
the huge optical depth of the neutral gas to these photons.  This is easily shown by equation 20, 
which for a largely neutral medium (i.e. $f_{\rm HI}$ $\simeq$ 1) yields a trapping radius on the 
order of 10$^{21}$ cm for the lowest accretion rates given by our solution (equation 12); as this is 
much larger than both $r_{\rm HII}$ and the Bondi radius, the maximum distance from which the 
protogalactic gas in can accrete onto the star.\footnote{The Bondi radius is roughly $r_{\rm Bondi}$ 
$\simeq$ 10$^{16}$ ($M_{\rm *}$/100 M$_{\odot}$) cm, assuming a sound speed in the 
protogalactic gas of 10 km s$^{-1}$, following \citet{wta08} and \citet{sbh10}.}  Therefore, we can 
also safely conclude that Ly$\alpha$ photons will not escape from the H~{\sc ii} region to affect the 
dynamics of the accretion flow at larger radii. 


We note that although most Ly$\alpha$ emission is trapped deep in the H II region, the thermal state of the gas 
may not be greatly affected because cooling via other atomic transitions in hydrogen still occurs 
\citep[see e.g][]{O01,raiter10,ssg10}.  At
the high densities within the H~{\sc ii} region atoms of both hydrogen
and helium are readily excited to higher energy ($n$ $>$ 1) states by both collisions
and absorption of photons; in turn, Lyman series photons
(principally Ly$\alpha$) are easily destroyed by ionizing
these excited atoms (see e.g. Osterbrock \& Ferland 2006). However, the radiative decay of
excited hydrogen atoms also results in significant emission of Balmer series
photons as well as two-photon emission from the H(2$s$) state, to
which the H~{\sc ii} is much less optically thick.  The net result is that the
energy in Lyman series photons is reprocessed largely into Balmer
series and two-photon continuum emission that escapes the H~{\sc ii} region 
and cools the gas \citep[see e.g.][]{ssg10}.  
Thus, photoionization heating of the gas can still be
balanced by efficient radiative cooling, and  even though much of the stellar radiation is reprocessed 
in the H~{\sc ii} region it still eventually emerges and may be observable (see $\S \, 4$).\footnote{To 
be precise, for a star with an effective temperature of 10$^5$ K, roughly 60 percent of its luminosity is 
emitted in ionizing photons, which are later reprocessed into nebular continuum and recombination 
emission, such as Ly$\alpha$.}  
It follows that the luminosity of the outgoing radiation is always the 
same as that of the star; since the star is taken to shine at the Eddington limit, this in turn implies 
that the force due to electron scattering indeed balances the force due to gravity everywhere in the 
highly ionized H~{\sc ii} region, as we assumed in our
calculation. This also suggests that the temperature of 
the gas is low enough to keep the flow supersonic so gas pressure cannot stop accretion, as we 
discuss in the Appendix.  

Overall, we conclude that it is the momentum imparted to the accreting
gas by ionizing photons, despite their being mostly converted 
to Ly$\alpha$ photons that are in turn converted largely to Balmer
series photons, that primarily regulates the growth of supermassive primordial stars.

\begin{figure*}[t]
  \centering
  \includegraphics[width=4.75in]{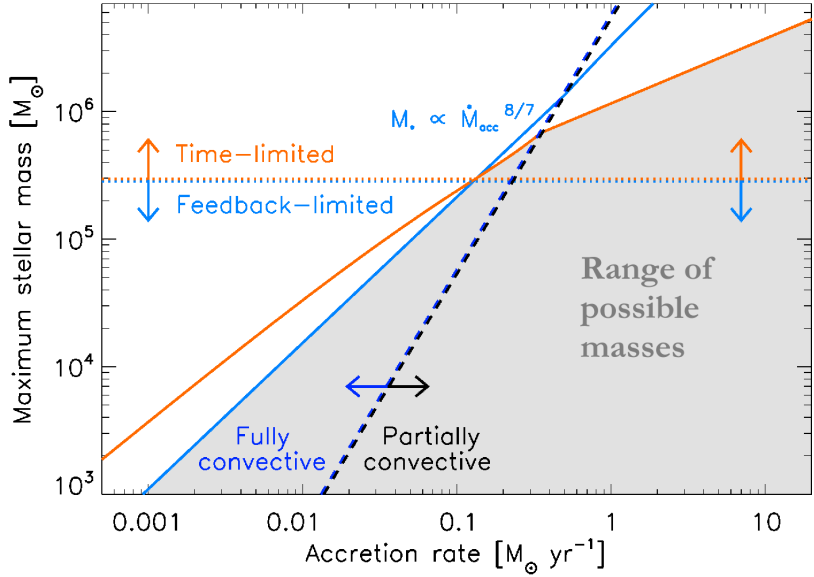}
  \caption{
The maximum mass to which a star can grow at a constant accretion rate $\dot{M}_{\rm acc}$.  
Above the maximum mass $M_{\rm *}$ $\propto$ $\dot{M}_{\rm acc}^{8/7}$ given by equation 15 ({\it blue line}), radiative feedback shuts 
off accretion because the H~{\sc ii} region of the star breaks out and prevents gas infall.  For accretion rates $\ga$ 10$^{-1}$ M$_{\odot}$ 
yr$^{-1}$, the maximum mass is set instead by the lifetime of the star ({\it red line}).  While at 
relatively low accretion rates the lifetime of an accreting massive primordial star is $\sim$ 4 
Myr, at high accretion rates the stellar lifetime can be shortened dramatically due to the 
super-Eddington luminosity at which nuclear fuel must be burned to support it during its growth.  
Also, at high accretion rates (right of the dashed curve), the star ceases to be fully convective 
and only the material in the convective core of the star is available for nuclear burning, further 
limiting the lifetime of the supermassive star.  The dotted horizontal line marks the mass ($\sim$ 3 $\times$ 10$^5$ M$_{
\odot}$) above which the final mass of a star is dictated by the limited time available for growth 
and below which it is governed by radiative feedback.  The shaded area denotes the range of 
possible stellar masses for accretion at a constant rate.
}
\end{figure*}

\vspace{0.1in}

\section{Time-Limited Accretion}

As noted by \citet{begel10}, the lifetime of a very massive primordial star burning nuclear 
fuel at the Eddington rate $L_{\rm Edd}$ and growing by accretion at a constant rate is $t_{\rm life}$ $\simeq$ 
4 Myr, twice the lifetime of non-accreting Pop III stars of similar mass.\footnote{This statement 
assumes that nuclear burning commences when the star has a mass much lower than its final 
mass.  However, we note that for Pop III stars with final masses of $\sim$ 100 M$_{\odot}$ 
nuclear burning may not begin until the star acquires a substantial fraction of its final mass 
\citep{op03}; if so, the lifetime of the star will be much closer to the $\sim$ 2 Myr expected for 
$\ga$ 100 M$_{\odot}$ Pop III stars of constant mass.}  However, the stellar lifetime is in fact 
shorter than this for two reasons.  First, nuclear fuel must be burned at a higher rate than the 
Eddington luminosity $L_{\rm Edd}$ in order to support the star
against its constantly increasing mass.  
Thus,  the star consumes fuel at the rate given by the sum of the
Eddington rate and the rate at which the binding energy of the star
increases (Begelman 2010):

\begin{equation}
L_{\rm nuc}  \simeq  L_{\rm Edd} + \frac{G M_{\rm *} \dot{M}_{\rm acc}}{r_{\rm *}} = L_{\rm Edd} \left(1 + \frac{r_{\rm trap}}{r_{\rm *}}  \right) \mbox{\ ,}
\end{equation}   
where $r_{\rm *}$ is the radius of the star (equation 14), $r_{\rm
  trap}$  $\equiv$ $\dot{M_{\rm acc}}$$\sigma_{\rm
  T}$/4$\pi$$m_{\rm H}$$c$   =  6 $\times$ 10$^{13}$ cm 
($\dot{M}_{\rm acc}$/M$_{\odot}$ yr$^{-1}$) is the radius at which radiation is 
trapped in the accretion flow by electron scattering, and $L_{\rm
  Edd}$ $\equiv$   4$\pi$$G$$M_{\rm *}$$m_{\rm H}$$c$/$\sigma_{\rm T}$   = 1.2 $\times$ 10$^{40}$ 
erg s$^{-1}$ ($M_{\rm *}$/100 M$_{\odot}$) is the Eddington luminosity
for a star of mass $M_{\rm *}$.  This effect of super-Eddington
nuclear burning contributes to the gradual turnover of the
stellar lifetime-limited
maximum stellar mass curve shown in red in Figure 3.

While equation (23) is valid when radiation can escape from the stellar
surface at $r_{\rm *}$ and the star is thus thermally relaxed, it
ceases to be when radiation is trapped in the accretion flow within a
radius $r_{\rm trap}$ above the star due to Thomson scattering.  
This occurs for accretion rates above $\dot{M}_{\rm acc, trap}$ $\simeq$ 4.3 $\times$ 10$^{-3}$ M$
_{\odot}$ yr$^{-1}$ ($M_{\rm *}$/100 M$_{\odot}$)$^{\frac{1}{2}}$.  As
radiation is trapped in the accretion flow, the accreting material can
not radiate its energy within $r_{\rm trap}$.  Upon reaching the
stellar surface it therefore deposits energy at a rate 
$GM_{\rm *}\dot{M_{\rm acc}}$/$r_{\rm trap}$, instead of at the higher rate 
$GM_{\rm *}\dot{M_{\rm acc}}$/$r_{\rm *}$ assumed in equation
(23).  Therefore, with this substitution is equation (23), at
accretion rates $\dot{M_{\rm acc}}$ $>$ $\dot{M}_{\rm acc, trap}$
the star consumes fuel as twice the Eddington rate.  This results 
in a decreased lifetime $t_{\rm life}$, which in turn leads
to the sharp turnover of the time-limited maximum stellar 
mass curve in Figure 3.

The second reason that the stellar lifetime is shortened is also
related to the trapping of radiation above the
surface of the star, as when this occurs the stellar envelope ceases
to be convective  \citep{begel10}.\footnote{In this 
event, the stellar photosphere also expands to $r_{\rm trap}$, as
previous authors have found in the case of accreting primordial protostars \citep{stahler86b,op03}.  Since the internal 
temperatures of primordial protostars are lower than on the main sequence, their opacities are 
somewhat higher than those due to Thomson scattering alone.  Consequently, their radii are found to swell 
to more than $r_{\rm trap}$ and perhaps terminate accretion, as
discussed in \citet{op03}.}  This results in the envelope material
never reaching the core and being available for fusion .  The fuel supply
of the star is thus limited to the mass within its convective core, whose fraction of the total 
mass is \citep[equation 26 of][]{begel10}    
\begin{equation}
\frac{M_{\rm conv}}{M_{\rm *}} \simeq 0.54 \left(\frac{r_{\rm trap}}{r_{\rm *}}\right)^{\frac{2}{3}} \left(\frac{\dot{M}_{\rm acc}}{{\rm M}_{\odot} \, {\rm yr}^{-1}} \right)^{-\frac{2}{3}} \left(\frac{M_{\rm *}}{10^6 \, {\rm M}_{\odot}} \right)^{\frac{1}{3}} \mbox{\ ,}
\end{equation}
where we have assumed a central temperature of 10$^8$ K.  
This second effect comes into play on the right hand 
side of the dashed line separating the fully convective regime from
the partially convective regime
(also separating the regimes in which $r_{\rm *}$ $>$ $r_{\rm trap}$
and $r_{\rm *}$ $<$ $r_{\rm trap}$).  

As shown in Figure 3, these two effects ultimately limit the final mass of the star to
$\sim$ 10$^6$ M$_{\odot}$ at the highest accretion rates ($\sim$ 1
M$_{\odot}$ yr$^{-1}$) found in cosmological simulations of the 
formation of supermassive stars via direct protogalactic collapse
(e.g. Wise et al. 2008; Shang et al. 2010; Johnson et al. 2011).
That said, we emphasize that our calculations have been done 
with significant simplifying assumptions, some of which we 
brielfy discuss in Section 5.  In particular, we have adopted a simple
model for the evolution of rapidly accreting supermassive stars;
however, to be fully confident in our results will require dedicated 
stellar evolution calculations that account for the continued growth of the star
over a large portion of its lifetime (Heger et al. in prep).  

Finally, we note that 'dark stars', which are powered by dark matter annihilation, could grow 
to considerably higher masses than those powered by fusion \citep[e.g.][]{freese08}.  This is 
because the cooler surface temperatures of such stars exert much less radiative feedback on 
accretion and dark matter fuel may last for a much longer time than nuclear fuel.  However, recent 
high resolution cosmological simulations show that primordial stars forming at the centers of dark 
matter halos may not remain there because of dynamical interactions, as required for the continual 
capture and annihilation of dark matter in their interiors \citep{stacyDS11,greif11}. Therefore, it may be 
that the final masses of dark stars are not so different from those expected for standard primordial 
stars \citep[see also][]{ripa10}.  If, however, dark stars do grow to be more massive they would 
exhibit spectral properties distinct from those of solely fusion-powered supermassive Pop III stars 
\citep{zack10,if11}, as we discuss next.

\begin{figure*}[t]
  \centering
  \includegraphics[width=4.75in]{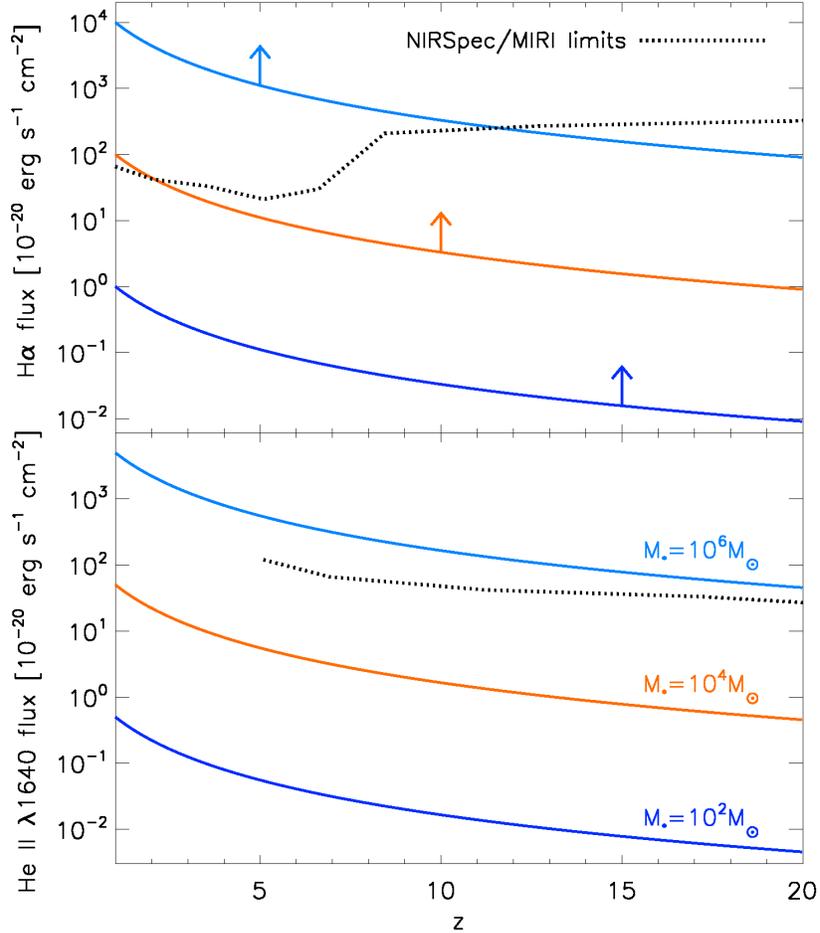}
  \caption{
The flux in H$\alpha$ (top panel) and He~{\sc ii} $\lambda$1640 (bottom panel) from accreting 
main sequence primordial stars, as a function of redshift $z$, for stellar masses of 10$^2$ ({\it 
dark blue}), 10$^4$ ({\it red}), and 10$^6$ M$_{\odot}$ ({\it light blue}).  The dotted black lines 
show flux limits for 3$\sigma$ detection of these lines in 10$^4$ second exposures with the 
NIRSpec and MIRI instruments on the JWST, operating at resolutions $R$ = 1000 and $R$ = 
1200-2400, respectively.  NIRSpec operates at wavelengths of $\simeq$ 1 to 5 $\mu$m, while 
MIRI covers the range from $\simeq$ 5 to 28 $\mu$m.  As discussed in $\S \, 4$, as we have 
not accounted for the reprocessing of trapped Ly$\alpha$ photons into Balmer series photons 
(including H$\alpha$), the H$\alpha$ fluxes shown here are lower limits, as indicated by the 
arrows.  While detection of H$\alpha$ from stars with masses of $\ga$ 10$^4$ M$_{\odot}$ 
may be possible out to very high redshift (e.g. $z$ $\ga$ 10), only accreting primordial stars with 
masses of at least 10$^5$ M$_{\odot}$ are likely to be detectable in He~{\sc ii} $\lambda$1640.
}
\end{figure*}

\section{Observational Signatures}

As supermassive stars are intense sources of radiation that could be detected by current and 
future surveys, we now examine their observable signatures.  The fact that radiative feedback 
in most cases cannot terminate accretion onto supermassive Pop III stars in collapsing 
protogalaxies implies that their H II regions will be confined deep in their host halos for most of 
their lives.  It follows that their ionizing radiation is largely reprocessed into nebular emission
instead of escaping the halo and reionizing the IGM.  As discussed in $\S \, 2.2$, because the accretion flow 
is optically thick to the Ly$\alpha$ photons, they cannot directly exit the halo and be observed.  These
photons are instead further reprocessed largely into Balmer series photons which do escape the H II 
region and halo and propagate into the IGM.  Therefore, one likely signature of rapidly accreting,
isolated supermassive Pop III stars in high-redshift protogalaxies is a strong Balmer line flux 
accompanied by a conspicuous lack of Ly$\alpha$ emission.  A detailed radiative transfer 
calculation is necessary to quantitatively predict luminosities for all the Balmer lines in hydrogen, 
but we can place lower limits on the H$\alpha$ flux, which is expected to be the dominant line. 

Following \citet{s02} and \citet{raiter10}, we compute the luminosity $L_{\rm H\alpha}$ in H$\alpha$ 
as a function of $Q_{\rm eff}$
as discussed in $\S \, 2.1$.  Relating this ionizing photon emission rate to stellar mass with equation 
13 yields $L_{\rm H\alpha}$ $\simeq$ 4 $\times$ 10$^{38}$ ($M_{\rm *}$/100 M$_{\odot}$) erg s$
^{-1}$; we again emphasize that this is a lower limit because we exclude reprocessing of Ly$\alpha$ 
into H$\alpha$, which may dramatically boost H$\alpha$ luminosities above those estimated here.  
The H$\alpha$ flux at redshift $z$ is
\begin{eqnarray}
f_{\rm H\alpha} &   =   &  \frac{L_{\rm H\alpha}}{4 \pi D_{\rm L}^2}  \nonumber \\ 
               & \sim & 10^{-20} \; {\rm erg} \; {\rm s}^{-1} \; {\rm cm}^{-2}  \left(\frac{L_{\rm H\alpha}}{10^{40} \; {\rm erg} \; {\rm s}^{-1}}\right) \left(\frac{1+z}{10}  \right)^{-2}  \mbox{ \, }
\end{eqnarray}
where $D_{\rm L}(z)$ is the luminosity distance to redshift $z$.  We plot this flux as a function of 
stellar mass and redshift in Figure 4, with the arrows on the H$\alpha$ curves signifying lower limits.

The hard spectrum of ionizing radiation from hot primordial stars also creates an He~{\sc iii} region 
from which a large luminosity in He~{\sc ii} $\lambda$1640 emission is expected \citep{ohr01,tgs01,
bkl01,s02,jlj09}.  Since the optical depth to this line is low, we also estimate how much of its flux exits 
the halo. For the large ionization rates and densities in our study, we can apply the standard model of 
\citet{raiter10} to compute the luminosity of this line, which we find to be $L_{\rm 1640}$ $\simeq$ 2 
$\times$ 10$^{38}$ ($M_{\rm *}$/100 M$_{\odot}$) erg s$^{-1}$, where we again have used Table 1 
of \citet{bkl01} for the number of He~{\sc ii}-ionizing photons produced as a function of stellar mass.  
Then, replacing $L_{\rm H\alpha}$ with $L_{\rm 1640}$ in equation 34, we derive the flux in the He
{\sc ii} $\lambda$ 1640 line, which we plot with $f_{H\alpha}$ in Figure 4 as a function of stellar mass 
and redshift.

Also shown in Figure 4 are detection limits for two instruments that will be on board the {\it James 
Webb Space Telescope} ({\it JWST}): the Near-Infrared Spectrograph (NIRSpec) and the Mid 
Infrared Instrument (MIRI).  The black dotted curves in Figure 4 denote the flux limits for 3$\sigma$ 
detection of the H$\alpha$ and He~{\sc ii} $\lambda$1640 lines in an exposure time of 10$^4$ s at 
resolutions $R$ = 1000 for NIRSpec and $R$ = 1200-2400 for MIRI (Gardner et al. 2006).\footnote{
JWST bandpass data can be found at www.stsci.edu/jwst/science/sensitivity.}  Because ionizing 
photon rates vary linearly with stellar mass, so do recombination line fluxes.  Consequently, only 
the more massive accreting primordial stars will be detectable by the \textit{JWST}.  Noting again 
that our H$\alpha$ line fluxes are lower limits, stars with masses $\ga$ 10$^5$ M$_{\odot}$ may 
be detectable in H$\alpha$ out to $z$ $\ga$ 10, while accreting stars with masses on the order of 
10$^4$ M$_{\odot}$ could perhaps be detected out to somewhat lower redshift.  Detection of the 
He~{\sc ii} $\lambda$1640 line will only be possible from stars with masses of at least 10$^5$ 
M$_{\odot}$ for sufficiently long exposure times.  

We note that the enhanced Balmer line emission from these objects may pose some difficulty to
their identification as supermassive Pop III stars.  As mentioned earlier, the large ratio of He~{\sc 
ii} $\lambda$1640 flux to hydrogen recombination line flux in Ly$\alpha$ or H$\alpha$ is unique 
to Pop III stars.  If, however, much of the Ly$\alpha$ emission is converted to H$\alpha$, the drop
in this ratio could mask the primordial nature of the star.  That said, the absence of any Ly$\alpha$ 
emission would still likely reveal the central object to be an accreting supermassive star.  Another  
attribute of rapidly accreting supermassive stars is strong continuum emission below the Lyman 
limit, which is the sum of the stellar continuum and the nebular continuum \citep[e.g.][]{raiter10}; 
the latter would be substantial due to the exceptionally high densities expected in the H~{\sc ii} 
regions of these stars.  Indeed, it is possible that the sum of the stellar and nebular emission could 
be detected, for example, in the Deep-Wide Survey to be carried out with the Near-Infrared Camera 
(NIRCam) on the JWST.  

An important obstacle to finding these objects is that their numbers could be small.  As discussed 
by previous authors \citep[][see also Agarwal et al. in prep]{bl03,dhl04,dijkstra08}, supermassive stars cannot be so abundant that 
the black holes they create exceed observed limits on the black hole mass density and x-ray 
background.  This suggests that finding the black holes may be easier than detecting their 
progenitors.  First, they could accrete material for at least 10$^8$ yr and be luminous 
for far longer than the stars that created them. Second, future x-ray 
missions such as the {\it Joint Astrophysics Nascent Universe Satellite} ({\it JANUS}) 
\citep{Roming08,Burrows10}, {\it LOBSTER} \citep{lobster}, {\it SVOM} \citep{svom}, and the {\it 
Energetic X-ray Imaging Survey Telescope} ({\it EXIST}) \citep{grind05} will perform all sky surveys 
with far greater coverage than the JWST.  However, if the nascent black hole does not have an 
accretion disk it may only emit weakly in x-rays at birth \citep[][but see also Komissarov \& Barkov 
2010]{fwh01,fh11,suwa11}.  If so, the SMBH seed does not become visible until it begins to accrete 
surrounding protogalactic gas \citep{hr93,km05,li07,vb10,jlj11}.
 
Supermassive stars could also be detected if they explode as luminous supernovae.  However,
previous studies have concluded that $\ga$ 10$^3$ M$_{\odot}$ Pop III stars collapse to black 
holes without an explosion \citep{fuller86,fh11,montero11} \citep[but see][]{ohk06}.  At 140 - 260 
M$_{\odot}$, however, pair-instability supernovae occur \citep{het03} and may be observable by 
future missions such as the \textit{JWST} \citep{wa05,sc05,wl05,jw11,kasen11}.  Finally, we note that much 
of the continuum and line emission from rapidly accreting supermassive stars will appear in the 
near infrared background (NIRB) today.  Although their contribution to the NIRB may be small if they 
are rare, it could be detected by missions such as the {\it Cosmic Infrared Background Experiment} 
\citep[CIBER; e.g.][]{bock06}, which is designed to find signatures of primordial galaxy formation at 
$z$ $\ga$ 10. 

\section{Discussion and Conclusions}

We find that the masses of the stellar seeds of SMBH forming from baryon collapse in early 
protogalaxies are primarily governed by their intense ionizing UV flux and their finite lifetimes.  For 
spherically-symmetric accretion at constant rates $\dot{M}_{\rm acc}$ $\la$ 0.1 M$_{\odot}$ yr$^{-1}$, 
the maximum mass the star can reach is governed by radiative feedback and is $M_{\rm *}$ $\simeq$ 
10$^{3}$ ($\dot{M}_{\rm acc}$/10$^{-3}$ M$_{\odot}$ yr$^{-1}$)$^{\frac{8}{7}}$ M$_{\odot}$.  At higher 
masses, the H II region breaks out to large radii and terminates accretion.  We have verified that other 
forms of feedback, such as gas pressure, radiation pressure from trapped line emission, 
photodissociation of H$^-$, and scattering of Ly$\alpha$ photons are much less effective at slowing 
accretion (see $\S \, 2.2$ and the Appendix).

At accretion rates above $\ga$ 0.1 M$_{\odot}$ yr$^{-1}$ the lifetime of the star, not radiative 
feedback, determines its final mass by limiting the time for which gas can accumulate on the star. 
Radiative feedback limits supermassive Pop III stars to final masses of $\sim$ 3 $\times$ 10$^5$ 
M$_{\odot}$ and time-limited accretion limits them to $\sim$ 10$^6$
M$_{\odot}$ at the highest accretion rates ($\sim$ 1 M$_{\odot}$ yr$^{-1}$) found in numerical
simulations of protogalactic collapse \citep{wta08,sbh10,jlj11}.

We caution that our analytical calculations do not account for all conceivable effects that could stem 
the growth of supermassive stars.  For instance, \citet{tm08} note that rotation of infalling gas leads to 
lower circumstellar densities and larger H~{\sc ii} regions with greater radiative feedback \citep[see 
also][]{hoyy11,stacy11}, implying lower final stellar masses.  Our spherically-symmetric calculation 
excludes rotation, so the mass limits we find for a given accretion rate (Fig. 3) are upper limits.  We do, 
however, find agreement with \citep{oi02}, who performed a similar calculation, although they only 
considered the growth of stars up to $\sim$ 10$^3$ M$_{\odot}$.  Furthermore, we also note that the 
spherical symmetry and constant accretion in our models do not address accretion that is episodic or
clumpy and self-shielded from ionizing radiation from the star \citep{wet08b,wet10,krum09}.  These processes 
could cause accretion to proceed at lower time-averaged rates than those predicted here 
\citep[but see][]{kuiper11}.

Another process that could truncate the growth of supermassive stars well before feedback 
or stellar lifetimes is the onset of a general relativistic instability in the core of the star that 
causes it to collapse when it becomes sufficiently massive \citep{chandra64}.  Such instabilities 
are predicted to set in once the star has grown to $\sim$ 10$^5$ M$_{\odot}$ \citep{iben63,
fowler64}, but stellar rotation could stabilize it against collapse up to much larger masses \citep{fowler66,bk98,
baum99}.  While our models ignore rotation, the accreting gas is likely to have some \citep[e.g.][]{col03};
if so, the supermassive star will inherit the angular momentum of the gas from which it formed 
and perhaps bypass the general relativistic instability.  Additional insight into 
the processes that limit the growth of supermassive stars will be gleaned from stellar evolution 
calculations accounting for the continual accretion of mass at high rates (Heger et al. in prep).  

If supermassive stars formed and grew to masses of $\ga$ 10$^5$ M$_{\odot}$ in the early 
universe, recombination emission from their H~{\sc ii} regions may be bright enough to be 
detected by future missions such as the \textit{JWST}.  In particular, Ly$\alpha$ photons trapped in 
the accretion flow are reprocessed into Balmer series photons that could escape into 
the IGM.  Consequently, the formation of these objects is accompanied by distinctive 
strong H$\alpha$ emission together with strong continuum and He~{\sc ii} $\lambda$1640 
emission, the latter arising from the hard spectrum of hot Pop III stars.  However, their 
black holes may be easier to discover in observational surveys, given the small numbers and 
brief lifetimes of their progenitors.  Indeed, these black holes may be the very ones that have 
already been found at the centers of massive galaxies and quasars at high redshifts. 

Current numerical simulations of SMBH seed formation in early protogalaxies are now at an 
impasse because the formation of the hydrostatic supermassive protostar restricts Courant 
times to values that are too short to evolve central flows for even one dynamical time (but see
Johnson et al. 2011 for an alternative approach using the sink particle technique).  Our results 
suggest that in the early and intermediate stages of the growth of the star its evolution is 
essentially decoupled from flows on even slightly larger spatial scales. Consequently, it should 
now be possible to retreat from the extreme spatial resolutions previously applied to the protostar 
and evolve flows at the center of the galaxy over enough dynamical times to capture its structure 
at the time of the death of the star and breakout of x-rays from the BH seed into the IGM.
Thus, it will soon be possible to witness the births of the first quasars in the universe with 
supercomputers.

\section*{Acknowledgements}
We gratefully acknowledge the support of the U.S. Department of Energy through the LANL/LDRD Program for this work.
We would like to thank Stirling Colgate, Dave Collins, Alex Heger, Kevin Honnell, Sadegh Khochfar,
Tsing-Wai Wong, and Hao Xu for helpful discussions.  This work also
benefited from the constructive comments of an anonymous reviewer.  
JLJ gratefully acknowledges the support 
of a LANL LDRD Director's Postdoctoral Fellowship at Los Alamos National Laboratory.  DJW 
acknowledges support from the Bruce and Astrid McWilliams Center for Cosmology at CMU.  
Work at LANL was done under the auspices of the National Nuclear Security Administration of 
the U.S. Department of Energy at Los Alamos National Laboratory under Contract No. 
DE-AC52-06NA25396.  


\bibliographystyle{apj}

\bibliography{refs}

\appendix

\section{Neglected feedback processes}


Here, we discuss several processes which have a negligible impact on our estimate of the
the final maximum stellar mass given by equation 15.

\subsection{He~{\sc ii} Photoionization}

To assess the relative importance of He~{\sc ii} photoionization in slowing the accretion flow in 
the He~{\sc iii} region, we follow the argument in Section 2.1 for He~{\sc i} photoionization.  For 
the ratio of He~{\sc ii} photoionization pressure to H~{\sc i} photoionization pressure, we have
\begin{equation}
\frac{p_{\rm HeII}}{p_{\rm HI}} \simeq \frac{n_{\rm He}}{n_{\rm H}} \frac{h\nu_{\rm HeII}}{h\nu_{\rm HI}} \frac{\alpha_{\rm B,HeII}}{\alpha_{\rm B,HI}} \simeq 1.1 \mbox{\ ,}
\end{equation}
where we have taken it that $h\nu_{\rm HeII}/h\nu_{\rm HI}$ $\simeq$ 54.4 eV / 29 eV $\simeq$ 
1.8, and that $\alpha_{\rm B,HeII} / \alpha_{\rm B,HI}$ $\simeq$ 6.4 \citep{osterbrock06}.
Therefore, within the He~{\sc iii} region, the radiation pressures due to He~{\sc ii} and H~{\sc i} 
photoionizations are comparable.  This is convenient because in the He~{\sc iii} region it is 
recombination emission from He~{\sc ii}, not stellar photons, which largely keeps hydrogen 
photoionized \citep{osterbrock06}.  Therefore, while this diffuse recombination emission does 
not transfer appreciable outward momentum to the gas, the stellar photons that ionize He~{\sc 
ii} transfer roughly the same momentum to the gas that they would in ionizing H~{\sc i}.  As a 
consequence, solving in detail for the pressure due to He~{\sc ii} photoionization would result 
in an almost identical solution to the one we have found by just treating the H~{\sc ii} region.

\subsection{H$^{ -}$ Photodetachment}

In order for very high accretion rates onto a supermassive star in a primordial protogalaxy to be 
realized, the H$_{\rm 2}$ fraction in the accreting protogalactic gas must be very low 
\citep{bl03,ln06,spaans06,osh08,wta08}.  Indeed, the $\ga 10^4$ K virial temperatures at the 
centers of 10$^7$ - 10$^8$ M$_{\odot}$ halos heavily suppress H$_2$ fractions, so it is a good 
approximation to take it that the opacity due to absorption of photons by H$_{\rm 2}$ is also low.  
However, this does not imply that H$^-$ fractions are negligible because it forms from reactants 
whose abundances are not suppressed by H$_{\rm 2}$-dissociating backgrounds or high virial 
temperatures:
\begin{equation}
{\rm H} + {\rm e}^- \to {\rm H}^- + \gamma  \mbox{\ ,}
\end{equation}
where $\gamma$ denotes the emission of a photon.  In principle, the absorption of photons with 
energies $\ge$ 0.75 eV by photodetachment of H$^-$ could be an important channel by which
momentum can be imparted to inflow.  To determine the rate at which momentum that could be 
transferred to the gas, it suffices to estimate the equilibrium rate of H$^-$ formation, since this is
also the rate at which H$^-$ will be destroyed and momentum will be acquired by the gas 
\citep[see][]{abet97,chuz07}.  Using the density profiles from our calculations, which assume that 
the gas falls toward the star at the free-fall velocity above $r_{\rm HII}$ (Fig. 2), we have $n(r)$ 
$\la$ 10$^{12}$ $(r/r_{\rm HII})^{-\frac{3}{2}}$ cm$^{-3}$.  Numerical simulations of protogalactic 
collapse predict central free electron fractions $f_{\rm e}$ $\sim$ 10$^{-4}$ \citep[e. g.][]{sbh10}.  
With these as upper limits along with the rate coefficient $k_{\rm H-}$ for H$^-$ formation, 
integration over the density profile of the halo yields the H$^-$ formation rate $Q_{\rm H-}$ 
outside the H~{\sc ii} region:
\begin{equation}
Q_{\rm H-} \la k_{\rm H-} \int^{r_{\rm outer}}_{r_{\rm HII}} 4 \pi r^2 f_{\rm e} n^2  dr \simeq 10^{44} {\rm s}^{-1} \mbox{\ ,}
\end{equation}
where $k_{\rm H-}$ $\sim$ 10$^{-14}$ cm$^{3}$ s$^{-1}$ \citep{wishart79,abet97}.  We  
assume that hydrogen is predominantly neutral outside the H~{\sc ii} region and neglect 
the small H$^-$ fractions that may form in the H~{\sc ii} region.  Finally, to ensure that we have 
found a strong upper limit, we integrate this profile out to $r_{\rm outer}$ = 10$^{21}$ cm, which 
is approximately the virial radius of the atomically-cooled halos at high redshift that host the 
supermassive stars we are studying.  Even allowing for the high average photon energy for 
photodetachment $h\nu_{\rm H-}$ $\sim$ 10 eV for supermassive Pop III stars, the maximum 
momentum that could be transferred to the gas per unit time is
\begin{equation}
\frac{h\nu_{\rm H-} Q_{\rm H-}}{c} \la 10^{22} \, {\rm \, g \, cm \, s^{-2}} \mbox{\ .}
\end{equation}
This is orders of magnitude smaller than the momentum of the accretion flow in equation 18, 
so H$^{-}$ destruction does not contribute to radiative feedback.

\subsection{Radiation Pressure from Trapped Ly$\alpha$ and Balmer Series Lines}

In Section 2.2, we raised the possibility that Ly$\alpha$ photons may slow down accretion 
by scattering from neutral atoms at the edge of the H~{\sc ii} region.  While we showed that 
this is unlikely to alter the flow, mostly because Ly$\alpha$ photons are confined to $r_{\rm 
Ly\alpha}$, radiation pressure due to nebular emission lines in optically thick gas may be 
sufficient to reduce accretion.  As also mentioned in Section 2.2, trapped Ly$\alpha$ 
photons may be largely converted to Balmer series photons, in some cases after being destroyed 
by absorption by excited hydrogen or helium \citep{osterbrock06}.  These Balmer photons can 
then propagate outward and efficiently cool the gas.  The high temperatures, densities, and 
Ly$\alpha$ photons trapped in the H~{\sc ii} region cause a large fraction of the hydrogen 
atoms to be excited to the $n$ = 2 state there.  From this state, either collisions or absorptions 
will further excite the atoms, which often later decay by emitting a Balmer series photon. 

To evaluate the degree to which radiation pressure from trapped line photons alters the 
dynamics of accretion, we first estimate the optical depth to Balmer photons both inside and 
outside the H~{\sc ii} region. Within $r_{\rm HII}$, the density of neutral hydrogen atoms can 
be estimated from equation 21 and assuming that  $n \propto r^{-2}$, as implied by the nearly 
constant velocity profile of the gas in the H~{\sc ii} region (Fig. 2).  The neutral hydrogen 
column density $N_{\rm H}$ through the H~{\sc ii} region is then
\begin{eqnarray}
N_{\rm H}(<r_{\rm HII}) & \simeq & \int^{r_{\rm HII}}_{r_{\rm *}} f_{\rm HI}(r) \, n(r) dr \nonumber \\ 
& = & f_{\rm HI}(r_{\rm HII}) \, n(r_{\rm HII}) \frac{r^2_{\rm HII}}{r_{\rm *}} \simeq 10^{17} \, {\rm cm}^{-2} \mbox{\ ,}
\end{eqnarray}
which, from the scaling in the second equation, can be shown to have only a weak dependence 
on the stellar mass for $M_{\rm *}$ = 10$^2$ - 10$^6$ M$_{\odot}$ for our solutions to the 
minimum accretion rate.  The cross sections for absorption of Balmer series photons by H in 
the $n$ = 2 state are 19, 3.5, and 1.3 $\times$ 10$^{-17}$ cm$^{2}$ for H$\alpha$, H$\beta$, 
and H$\gamma$, respectively.  The higher energy Balmer series cross sections are all below 
these values.  With these cross sections and $N_{\rm H}(<r_{\rm HII})$, we can express the 
optical depth through the H~{\sc ii} region solely as a function of the relative populations $N_
{\rm 2}$/$N_{\rm 1}$ of the $n$ = 1 and $n$ = 2 levels of neutral hydrogen in the H~{\sc ii}
region.  We find that even for $N_{\rm 2}$/$N_{\rm 1}$ as high as $\sim$ 0.05, the H~{\sc 
ii} region is optically thin to all Balmer series photons, while the largest optical depth possible 
for H$\alpha$ is $\tau_{\rm H\alpha}$ $\simeq$ 20.  However, this is still far below the optical 
depth for Ly$\alpha$ photons, and Balmer series photons will eventually leak out of the H~{\sc 
ii} region, although in some cases only after a number of scatterings.  Therefore, while 
Ly$\alpha$ photons in general will not escape the H~{\sc ii} region, Balmer series photons will 
escape and allow the gas to radiatively cool, as we noted in Section 2.2.

Next, we consider the optical depth to Balmer series lines outside the H~{\sc ii} region.  In 
this region, we set $f_{\rm HI}$ = 1 and adopt a free-fall density profile $n$ $\propto$ $r^{-\frac
{3}{2}}$ for the gas, as in Section 2.1.  The neutral hydrogen column density beyond $r_{\rm HII}$ 
is then 
\begin{eqnarray}
N_{\rm H}(>r_{\rm HII}) & \simeq & \int^{\infty}_{r_{\rm HII}} n(r) dr \nonumber \\ 
& = &  2 n(r_{\rm HII}) r_{\rm HII} \simeq 10^{24} \, {\rm cm}^{-2} \mbox{\ ,}
\end{eqnarray}
which again is not strongly dependent on stellar mass.

Because the gas outside the H II region is well below the critical density at which excited levels 
in hydrogen would be populated and kept in equilibrium by collisions (and because Ly$\alpha$ 
photons are mostly trapped in the H~{\sc ii} region and cannot excite neutral hydrogen beyond
$r_{\rm HII}$), the relative population of the $n$ = 2 level of hydrogen is expected to be very small.  
In this case, \citet{ssg10} find that in general $N_{\rm 2}$/$N_{\rm 1}$ $\la$ 10$^{-10}$.  With this 
as an upper limit and the cross sections for Balmer series photon absorption above, we find that 
the optical depths to Balmer series lines are $\tau_{\rm H\alpha}$ $\sim$ 4 $\times$ 10$^{-2}$, 
$\tau_{\rm H\beta}$ $\sim$ 8 $\times$ 10$^{-3}$, and $\tau_{\rm H\gamma}$ $\sim$ 3 $\times$ 
10$^{-3}$.  At such low optical depths, Balmer series photons will propagate largely unimpeded 
through the inflow and, most likely, exit the halo with minimal effect on accretion.  As these 
photons also move freely through the intergalactic medium (IGM), accreting supermassive stars 
probably have a distinctive observational signature, as we discussed in $\S \, 4$.

Thus, because outside the H~{\sc ii} region the optical depth to Balmer series photons is small 
and Ly$\alpha$ photons cannot propagate across the H~{\sc ii} region boundary, we conclude 
that radiation pressure due to trapped emission lines will not be appreciable beyond $r_{\rm HII}$.  
However, it may be that this pressure is substantial within the H II region.  To estimate its magnitude,
we compare it to the ram pressure of the accretion flow.  For the latter, we have 
\begin{eqnarray}
P_{\rm ram} & = & n \mu m_{\rm H} v^2 \nonumber \\
& \simeq & 50 \left(\frac{\dot{M}_{\rm acc}}{{\rm M_{\odot}} \, {\rm yr^{-1}}} \right) \left(\frac{r}{10^{15} \, {\rm cm}} \right)^{-2} \left(\frac{v}{100 \, {\rm km} \, {\rm s}^{-1}} \right) {\rm dyn} \, {\rm cm}^{-2}  \mbox{\ ,} 
\end{eqnarray}
where we have used equation 1 to express $n$ in terms of $v$ and ${\dot{M}}_{acc}$.  We 
use the prescription of \citet{bd89} for the radiation pressure \citep[see also][]{ef86}:
\begin{equation}
P_{\rm line} = \left(\frac{4\pi}{9c} \right) \left(\frac{2h\nu_{\rm tran}^3}{c^2} \right) \frac{N_{\rm i+1}}{N_{\rm i}} \Delta \nu_{\rm tran} \mbox{\ ,}
\end{equation}
where $N_{\rm i+1}$/$N_{\rm i}$ is the ratio of the upper and lower level populations for the 
given transition, whose frequency is $\nu_{\rm tran}$.  The line width $\Delta \nu_{\rm tran}$ 
$\sim$ 2 $\times$ 10$^{11}$ ($T$/10$^4$ K)$^{\frac{1}{2}}$ (ln $\tau$)$^{\frac{1}{2}}$ s$^{-1}$, 
where $\tau$ is the optical depth of the line.  

Assuming optical depths $\tau_{\rm Ly\alpha}$ = 10$^4$ and $\tau_{\rm Balmer}$ = 20, both 
of which are rough upper limits in the H~{\sc ii} region, and $N_{\rm i+1}$/$N_{\rm i}$ = 1 in 
both cases, we find strong upper limits of $\sim$ 6 dyn cm$^{-2}$ and 2 $\times$ 10$^{-2}$ 
dyn cm$^{-2}$ for the Ly$\alpha$ and Balmer line radiation pressures, respectively.  
Comparing these pressures to $P_{ram}$ at $r_{\rm HII}$, which is a lower limit for the H~{\sc 
ii} region due to its strong dependence on $r$ in equation A7, we find that it is always at least 
a factor of $\sim$ 4 larger than radiation pressure from optically thick lines.  Therefore, lines
will not play a large role in slowing accretion in the H II region.

\subsection{Gas Pressure}

Because infall is highly supersonic, gas pressure cannot decelerate the gas \citep{oi02}.  
We have verified that this holds even when a gas pressure term is included in the equation 
of motion.  We find that only for extremely high sound speeds corresponding to temperatures 
of at least 10$^6$ K would gas pressure begin to impact the dynamics of the accretion flow.  
However, temperatures this high are not found in numerical simulations of protogalactic 
collapse \citep{wta08,rh09,sbh10,jlj11} or in H~{\sc ii} regions because the Balmer 
thermostat limits ionized gas temperatures to at most a few $\times$ 10$^4$ K.

\subsection{Accretion Luminosity}

Our conclusions regarding the maximum mass of the supermassive star rest on the assumption 
that the critical accretion rate for a given stellar mass (equation 12) separates two regimes: 
below this rate, accretion is suppressed by radiative feedback and above this rate accretion 
proceeds.  As shown in Section 2.1, it is clear that accretion is prevented for inflow rates 
below the critical rate because the ionization front breaks out to infinity and photoionization 
pressure halts infall at all radii.  For accretion rates above the critical rate, the flow will arrive at 
the star with a velocity $v(r_{\rm *})$ $>$ 0.  This implies that the accreting material will have 
some kinetic energy that must be dissipated at the stellar surface, some portion of which will be
radiation. This radiation could slow accretion within the H~{\sc ii} region by electron scattering 
or photoionizations.  

Let us assume that all of the kinetic energy of the flow is converted to radiation that propagates
outward on impact with the star at some velocity $v(r_{\rm *})$.  The luminosity thus generated is 
\begin{equation}
L_{\rm acc} = \frac{\dot{M}_{\rm acc}v^2}{2} \mbox{\ ,} 
\end{equation}
which yields a total momentum in photons of $L_{\rm acc}$/c = $\dot{M}_{\rm acc}$$v^2$/2$c$.  
Comparing this to the momentum of the infalling gas, $\dot{M}_{\rm acc}$$v$, we see that the 
momentum in the accretion luminosity is a factor $v$/2$c$ lower than that of the gas. Therefore, 
accretion radiation cannot halt the flow in the steady-state approximation.  However, bouts of 
massive accretion could in principle generate enough radiation to slow infall at larger radii 
where gas has a lower momentum, like the episodic accretion onto black holes in the early 
universe discussed by \citet{milos09}.

We note that an additional effect of accretion at very high rates (much higher than those in 
equation 12) is the trapping of radiation by electron scattering or other sources of opacity in 
the flow \citep[e.g.][]{begel78,begel10,wl11}.  As discussed by \citet[][see also Stahler et al. 1986]{op03} 
in the context of primordial protostellar accretion, radiation trapping can lead to rapid 
expansion of the stellar photosphere that halts the growth of the star.  We note, however, 
that such expansion also causes the effective temperature of the star to fall with the increase 
in surface area of the photosphere\footnote{At constant luminosity, the effective temperature 
at the stellar photosphere scales as $T_{\rm eff}$ $\propto$ r$^{-\frac{1}{2}}_{\rm *}$}, which in 
turn leads to a softening of the emitted radiation and a lower ionizing photon emission rate $Q$.  
This, and the fact that accretion rates greater than those in equation 12 result in smaller H II 
regions, suggests that the flow cannot be stopped, at least not by photoionization pressure. 

\end{document}